\documentclass[journal]{IEEEtran}
\usepackage{cite}

\ifCLASSINFOpdf 
\usepackage[pdftex]{graphicx}
\else
\usepackage[dvips]{graphicx}
\fi

\usepackage[cmex10]{amsmath}
\usepackage{algorithmic}
\usepackage[ruled,norelsize]{algorithm2e}

\usepackage{array}

\usepackage{mdwmath}
\usepackage{mdwtab}
\usepackage{todonotes}

\usepackage{eqparbox}

\ifCLASSOPTIONcompsoc
\usepackage[caption=false,font=normalsize,labelfont=sf,textfont=sf]{subfig}
\else
\usepackage[caption=false,font=footnotesize]{subfig}
\fi

\usepackage{fixltx2e}
\usepackage{url}
\usepackage{amsfonts}
\usepackage{amssymb}
\usepackage{amsthm}
\usepackage{bm}
\usepackage[ruled]{algorithm2e}
\usepackage[utf8]{inputenc}
\usepackage{booktabs}
\usepackage{url}
\usepackage{cite}
\usepackage{flushend}
\usepackage{epstopdf}
\usepackage{multirow}

\newcounter{tempEquationCounter} 
\newcounter{thisEquationNumber}

\begin{document}
%
\title{Private 5G: The Future of Industrial Wireless}


\author{
Adnan~Aijaz,~\IEEEmembership{Senior~Member,~IEEE}
        \vspace{-1.0em}
\thanks{The author is with the Bristol Research and Innovation Laboratory, Toshiba Europe Ltd., Bristol, BS1 4ND, U.K. E-mail: adnan.aijaz@toshiba-bril.com}}
\markboth{Accepted for Publication in IEEE IEM}%
{Shell \MakeLowercase{\textit{et al.}}: Bare Demo of IEEEtran.cls for Journals}
%


\maketitle
\begin{abstract}
\boldmath
High-performance wireless communication is crucial in  digital transformation of industrial systems which is driven by Industry 4.0 and the Industrial Internet initiatives. \textcolor{black}{Among the candidate industrial wireless technologies, 5G (cellular/mobile) holds significant potential. Operation of private (non-public) 5G networks in industrial environments is promising to fully unleash this potential.} This article provides a technical overview of private 5G networks.  It introduces the concept and functional architecture of private 5G  while highlighting the  key benefits and  industrial use-cases. It explores  spectrum opportunities for  private 5G networks. It also discusses  design aspects of private 5G along with the key challenges. Finally, it explores the emerging standardization and open innovation ecosystem for private 5G. 
\end{abstract}
\begin{IEEEkeywords}
5G, industrial wireless, Industry 4.0, non-public networks, private networks, TSN, uRLLC, wireless control.
\end{IEEEkeywords}

%
\IEEEpeerreviewmaketitle

\section{Introduction}
\IEEEPARstart{T}{he} emergence of industrial networking \cite{marshall_ind} is one of the most notable developments in the history of industrial communication. Industrial networks provide various benefits in addition to overcoming the limitations of conventional point-to-point wired systems \cite{emer_ind_net}. \textcolor{black}{Industrial networks exist across a range of industrial domains including manufacturing,  oil and gas, power generation/distribution, mining, and chemical processing. Such networks  differ quite significantly from traditional enterprise/consumer networks in terms of service requirements. }

The information transmitted over industrial networks can be broadly categorized into control and monitoring (diagnostic). The control information is typically exchanged between controllers and industrial devices (sensors, actuators, etc.). It has strong deterministic real-time requirements. The monitory information is the sensory information which is used to monitor the status of industrial equipment. Typically, it is not acted upon in real-time.  The emerging Industry 4.0 \cite{ind4} (or the wider Industrial Internet \cite{ind_internet}) initiative aims at enhancing the efficiency, productivity, flexibility and versatility of legacy industrial systems. With the advent of Industry 4.0, several new industrial applications, with challenging and diverse requirements, have emerged that connect people, objects, processes and systems in real-time \cite{TI_PIEEE}. \textcolor{black}{\tablename~\ref{t1} summarizes the key connectivity requirements\footnote{\textcolor{black}{The term ``scalability'' in \tablename~\ref{t1} refers to the number of nodes/devices that need to be supported while meeting the latency and reliability requirements. }} of different industrial applications}.

\textcolor{black}{Wireless technologies provide a number of benefits for industrial communication such as greater flexibility of connecting machines and devices, reduction in installation and maintenance costs, built-in support for mobility, and reduced exposure of personnel to hazardous situations \cite{wless_ind}}. Wireless technologies are also expected to play a key role in realizing the vision of Industry 4.0. In state-of-the-art industrial systems, wireless technologies are mainly used for monitoring applications. The use of wireless for control applications is still at a nascent stage. This is due to the fact that most existing wireless technologies fall short of meeting the stringent requirement of control applications which pose significant challenges to the communication network in terms of determinism, service availability, high reliability, and low latency. Consequently, control applications are realized through wired technologies like fieldbus systems \cite{fieldbus} and industrial Ethernet \cite{ind_ethernet} which are costly and inflexible.

Industrial networks have evolved over three distinct generations \cite{gen_field_net}. A common trend across all generations has been the development of dedicated wired as well as wireless solutions  with little emphasis on interoperability \cite{fut_ind_comm}. As a result, industrial community's quest for a \emph{single connectivity solution} still remains unfulfilled. 
However, the industrial connectivity landscape is expected to change dramatically with the emergence of fifth generation (5G) cellular/mobile networks. Unlike previous  generations, 5G is expected to enable a range of use-cases across vertical industries. The design requirements of 5G coincide with some of the most demanding industrial control applications. \textcolor{black}{The most remarkable feature of 5G is its service-oriented approach, which unlike the one-size-fits-all approach of previous generations, provides the much-needed design flexibility. This is not offered by existing industrial wireless technologies (based on IEEE 802.15.4, Bluetooth, Wi-Fi, etc.)  that can only cater for the requirements of specific applications. }

The potential of 5G for industrial communication is widely recognized. \textcolor{black}{However, to fully unleash this potential, operation of \emph{private} 5G networks in industrial environments is important.} Private 5G offers dedicated coverage, exclusive use of resources and the opportunity for a customized service for industrial use-cases. Most importantly, private 5G  offers complete control over every aspect of the network. 

To this end, this article aims to provide a technical overview of private 5G networks. It begins by motivating the need for 5G for industrial communication. After that, it provides an overview of private 5G networks highlighting  key benefits, functional architecture aspects and prominent industrial use-cases. This is followed by a discussion on spectrum opportunities for private 5G deployments,  the key design aspects including network slicing techniques and integration with time-sensitive networking (TSN), and the emerging standardization and open innovation ecosystem for private 5G. The paper is concluded with  some key insights.

	\begin{table*}[]
\centering
\caption{Connectivity Requirements of Key Industrial Applications (based on \cite{3gpp.22.804} and \cite{TI_PIEEE}) }
\begin{tabular}{lcccc}
\cline{1-5}
				\toprule
\textbf{Application}                                                    &  \textbf{\textcolor{black}{Reliability}} & \textbf{Latency / Cycle Time                                                   } & \begin{tabular}[c]{@{}c@{}}\textbf{Data Rate} \end{tabular} & \textbf{\textcolor{black}{Scalability} } \\ \cline{1-5} \midrule
\multicolumn{5}{c}{\textbf{Conventional Industrial Applications}}\\
\midrule
\begin{tabular}[c]{@{}c@{}}Monitoring\end{tabular} & \(\geq 99.9\%\)           & \begin{tabular}[c]{@{}c@{}}\(50\) ms -- \(100\) ms \end{tabular} & \(0.1\) Mbps -- \(0.5\) Mbps & \(100\) -- \(1000\) nodes
\\ 
\begin{tabular}[c]{@{}c@{}}Safety Control\end{tabular} & \(\geq 99.999\%\)        & \begin{tabular}[c]{@{}c@{}} \(5\) ms -- \(10\) ms \end{tabular} & \(0.5\) Mbps -- \(1\) Mbps & \(10\) -- \(20\) nodes
\\
\begin{tabular}[c]{@{}c@{}}Closed-loop Control\end{tabular} & \(\geq 99.999\%\)          & \begin{tabular}[c]{@{}c@{}}\(2\) ms -- \(10\) ms \end{tabular} & \(1\) Mbps -- \(5\) Mbps & \(100\) -- \(150\) nodes
\\
\begin{tabular}[c]{@{}c@{}}Motion Control \end{tabular} & \(\geq 99.9999\%\)           & \begin{tabular}[c]{@{}c@{}}\(0.5\) ms -- \(2\) ms \end{tabular} & \(1\) Mbps -- \(5\) Mbps & \(10\) -- \(50\) nodes
\\ \cline{1-5} 
\midrule
\multicolumn{5}{c}{\textbf{Emerging Industrial Applications}}\\
\midrule 
\begin{tabular}[c]{@{}c@{}}Mobile Workforce\end{tabular} & \(\geq 99.999\%\)          & \begin{tabular}[c]{@{}c@{}}\(5\) ms -- \(10\) ms \end{tabular} & \(10\) Mbps -- \(50\) Mbps & \(50\) -- \(100\) nodes
\\ 
\begin{tabular}[c]{@{}c@{}}Augmented Reality\end{tabular} & \(\geq 99.99\%\)         & \begin{tabular}[c]{@{}c@{}}\(5\) ms -- \(10\) ms \end{tabular} & \(500\) Mbps -- \(1000\) Mbps & \(10\) -- \(20\) nodes
\\
\begin{tabular}[c]{@{}c@{}}Remote Maintenance\end{tabular} & \(\geq 99.99\%\)          & \begin{tabular}[c]{@{}c@{}}\(20\) ms -- \(50\) ms \end{tabular} & \(1\) Mbps -- \(2\) Mbps & \(500\) -- \(1000\) nodes
\\
\begin{tabular}[c]{@{}c@{}}Remote Operation\end{tabular} & \(\geq 99.999\%\)          & \begin{tabular}[c]{@{}c@{}}\(2\) ms -- \(10\) ms \end{tabular} & \(100\) Mbps -- \(200\) Mbps & \(1\) -- \(5\) nodes
\\ \cline{1-5} 
\end{tabular}
\label{t1}
\end{table*}

\section{Why 5G for Industrial Communication?}
5G is particularly attractive for industrial communication due to a range of factors.

\textbf{Unified wireless interface} -- 5G offers a unified wireless interface to support the diverse requirements of different industrial applications. It has been designed for  three main service categories \cite{ITU_2083}: enhanced mobile broadband (eMBB) with peak data rate of up to \(10\) Gbps, massive machine type communication (mMTC) with connection density of up to \(100\) nodes per sq. meters and ultra-reliable low latency communications (uRLLC) providing \(1\) millisecond (ms) user-plane latency with \(>99.999\%\) reliability.  \textcolor{black}{Such a capability is not offered by any existing industrial wireless technology.}


\textbf{Guaranteed Quality-of-Service (QoS)} -- Unlike traditional industrial wireless technologies based on Wi-Fi or Bluetooth, 5G provides guaranteed QoS for critical industrial applications. 

\textbf{Mobility} -- Mobile platforms like automated guided vehicles  and mobile robots are a key feature of various emerging and future industrial systems \cite{TI_PIEEE}. 5G provides built-in support for handling mobility in industrial environments.

\textbf{Security} -- 5G brings a proven and tested security technology to the industrial world which has been deployed in cellular networks worldwide. 

\textbf{Positioning} -- Positioning capabilities, which have received significant attention in previous cellular generations, are under consideration for 5G. An early assessment of  positioning requirements reveals that the targets are quite high (e.g., accuracy within \(10\) cm with latency of \(<15\) ms) \cite{3gpp_22_872}. Such high-accuracy positioning capabilities would be crucial for many emerging industrial applications. 

\textcolor{black}{The recent IEEE 802.11ax (Wi-Fi 6) technology is a potential competitor to 5G. It provides much lower latency as compared to previous Wi-Fi generations, which makes it attractive for industrial applications \cite{aijaz_IES}. However, 5G has a distinct advantage over Wi-Fi 6 for industrial communication due to a number of factors, such as flexible air-interface design, native support for uRLLC and mobility, QoS guarantees, and better scalability.  }

%
%
%

%
%
%
%
%

\begin{figure*}
\centering
\includegraphics[scale=0.64]{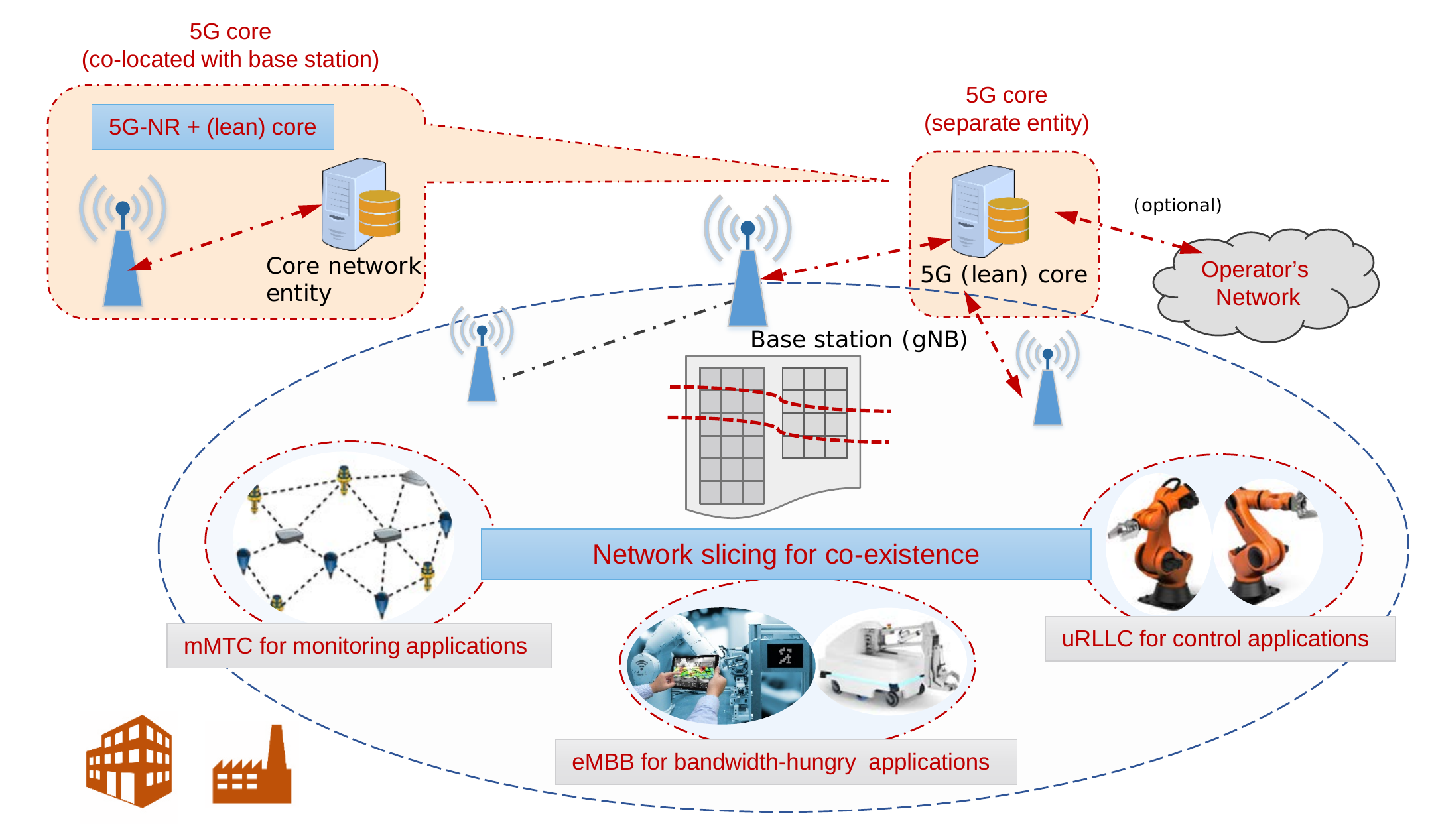}
\caption{The concept of a private 5G network for industrial communication based on network slicing for mMTC, eMBB and uRLLC applications.  }
\label{c_pvt5g}
\end{figure*}

\section{Private 5G Networks}
\subsection{Concept}
By definition, a private 5G network (illustrated in Fig. \ref{c_pvt5g})  is a local area network\footnote{It is emphasized that not every local 5G network is a private network.}, based on 5G new radio (NR) technology, for dedicated wireless connectivity in a specific region. The radio access network (RAN) part of the private network comprises a single or  multiple base stations. The base stations can scale as per the capacity and coverage requirements. The core network part of a private 5G network is rather lean as compared to its public counterpart. Physically, it can be a separate entity in the network or co-located  with the base station in the same box.  A private 5G network can be deployed for a specific industrial application or for multiple industrial applications with diverse requirements. \textcolor{black}{Private networks are also referred to as \emph{non-public networks} (NPNs) in 3GPP. }

\subsection{What Private 5G Offers?}
The unique aspect of private 5G is that it empowers industrial players to run their own local networks with dedicated equipment and settings. 

\textbf{Dedicated Coverage} -- Private 5G networks offer dedicated coverage at a facility or location. This is particularly important for industrial sites which are often located at remote locations where public networks do not exist or indoor coverage is not robust. Such dedicated coverage is crucial to achieving very high availability for industrial operations. 

\textbf{Exclusive Capacity} -- A private 5G network makes exclusive use of the available capacity. There is no contention from other network users as on a public network.

\textbf{Intrinsic Control} -- A private 5G network offers the possibility of complete control to its owner -- something which is not possible on the public networks. Private operators  can deploy their own security policies to authorize users, prioritize traffic, and most importantly, to ensure that sensitive data does not leave the premises.

\textbf{Customized Service} -- A private 5G network can be customized as per the requirements of specific industrial applications. Such customization is not possible on a public network.  Moreover, a private 5G network can be efficiently shared among multiple industrial applications. 

\textbf{Dependable Communication} -- 
The dedicated nature of private 5G networks coupled with customized service, intrinsic control and uRLLC capabilities provide dependable industrial wireless communication. 


\subsection{Functional Architecture}
\begin{figure*}
\centering
\includegraphics[scale=0.45]{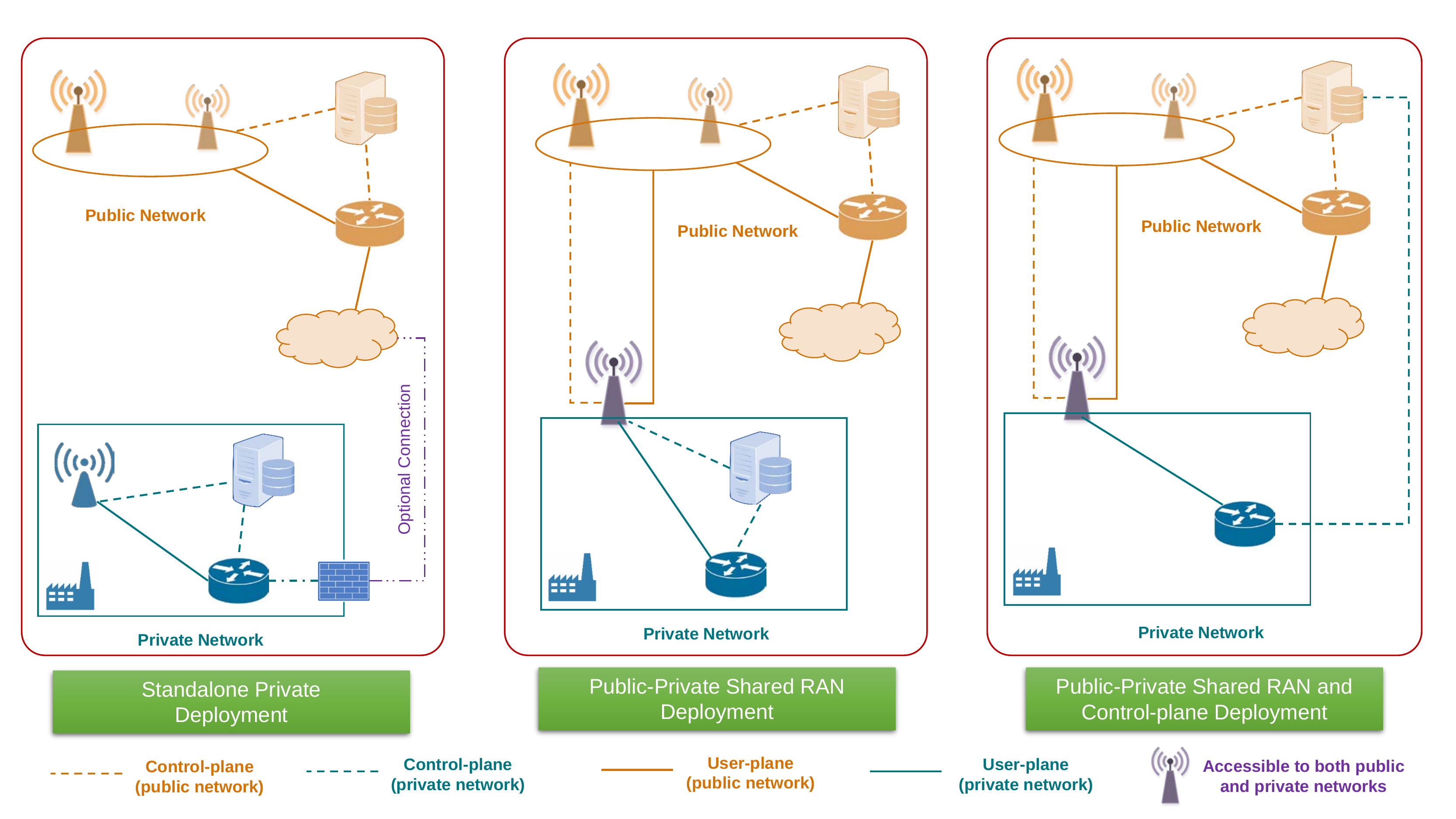}
\caption{Functional architecture of private 5G networks: standalone private deployment, public-private shared RAN deployment and public-private shared RAN and control-plane deployment (adapted from \cite{5G_ACIA_NPN}). }
\label{func_arch}
\end{figure*}

From a functional perspective, private 5G networks can be deployed in the following ways, which are also illustrated in Fig. \ref{func_arch}.

\begin{itemize}
\item \textbf{Standalone Deployment} -- In this case, the private 5G network is deployed as a standalone independent network. \textcolor{black}{The standalone private network (or standalone NPN) is completely separate from the public network and all the data flows and network functions (user-plane as well as control-plane) take place inside the premises of the industrial site (e.g., a warehouse or a factory).} There is an option of connecting with the public network through a firewall if access to public network services is required. 

\item \textbf{Public-Private Shared RAN Deployment} -- In this case, the private 5G network shares the RAN with the public network; however, all network functions remain separate. Moreover, all data flows of the private 5G network are confined to the premises of the industrial site. The private network has its own identity; however, there is a RAN sharing agreement with the public network. The 3GPP multi-operator core network (MOCN) model \cite{3gpp_23_251} for RAN sharing is the key enabler for realizing such deployments. 

\item \textbf{Shared RAN and Control-plane Deployment} -- In this case, the public and private networks share part of the RAN. Moreover, control-plane network functions are always handled on the public network. \textcolor{black}{Such a deployment is realized through network slicing techniques (described later)}. All the private network data flows are within the premises of the industrial site. The private network and the public network have separate slice identifiers. The private network users are actually the subscribers of the public network. Hence, roaming on the public network and access to its service is rather straightforward.

\end{itemize}

In addition to the aforementioned deployment options, the private network can  be deployed over a neutral host infrastructure. However, in this scenario, the private network data flows are not necessarily confined to the the industrial premises. 

\begin{figure}
\centering
\includegraphics[scale=0.4]{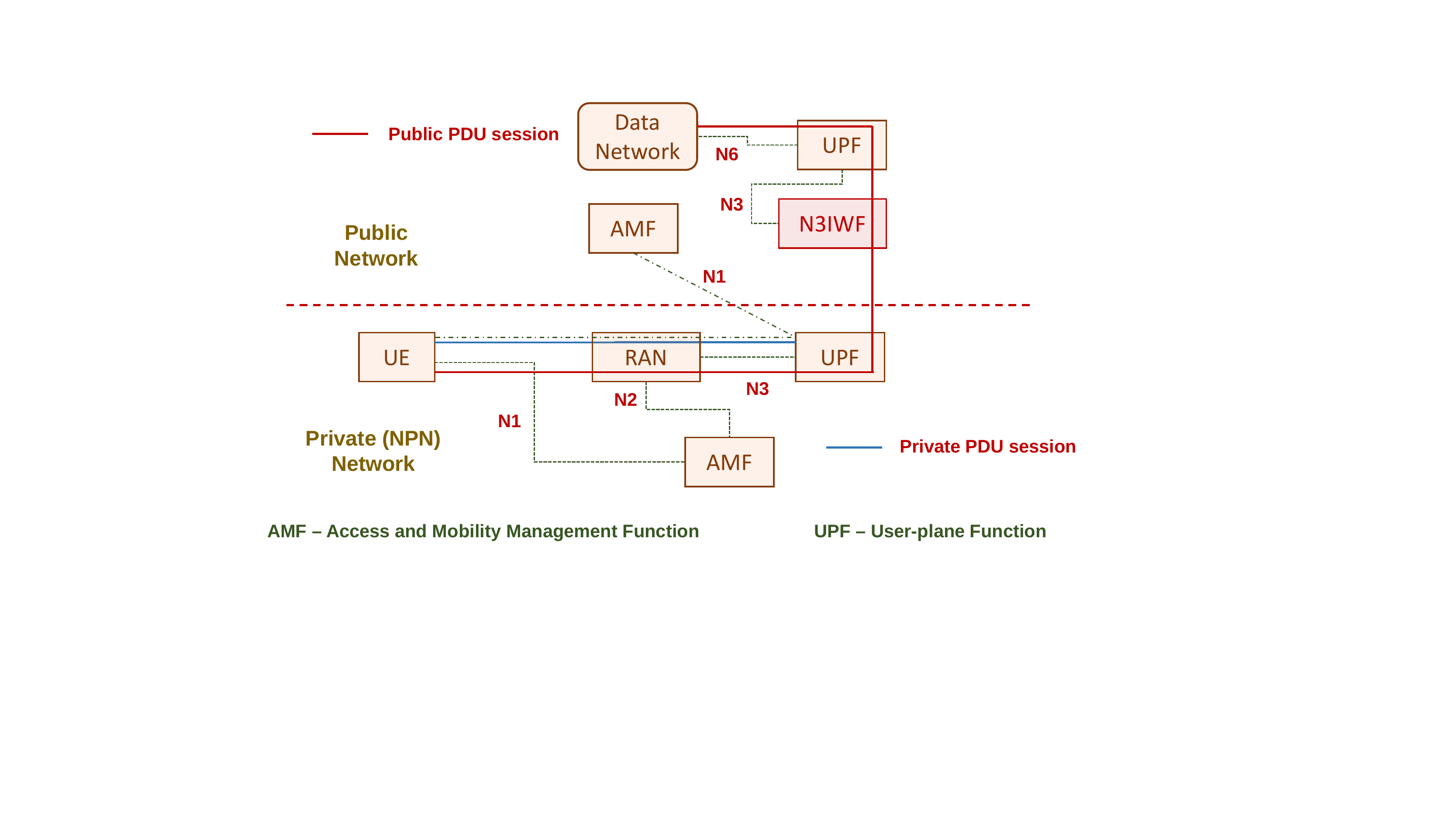}
\caption{Access to public network services via the private network.}
\label{sc_pvt}
\end{figure}

\textcolor{black}{3GPP has proposed two main solutions for service continuity between private and public networks \cite{3gpp_23_734}. The first solution is based on dual-radio capabilities. A user can be simultaneously registered and in connected state in both private and public networks. The second solution, illustrated in Fig. \ref{sc_pvt}, is based on non-3GPP interworking function (N3IWF). For access to public network services via a private network, a user obtains IP connectivity via the private network, discovers an N3IWF provided by the public network and establishes connectivity to the public network via the N3IWF.      }

\subsection{Key Industrial Use-cases}
Private 5G networks can be deployed for a wide range of use-cases across different industrial domains. \textcolor{black}{According to a recent report   \cite{harbor_pvt}, private cellular networks for industrial markets are expected to generate \(\$70\) billion in revenue by 2023.} Some of the key use-cases of private 5G networks are discussed as follows.

\textbf{Industrial Automation} --
Private 5G networks can be deployed inside factories and manufacturing sites to fulfil the stringent connectivity requirements of industrial automation networks. Private 5G can support field-level communication between industrial controllers and field devices (sensors, actuators, etc.) as well as communication between industrial controllers. Such communication is characterized by the requirements of very low latency and very high reliability, and it is currently realized through wired technologies. Private 5G offers a promising opportunity for cable replacement in industrial automation networks. Private 5G also provides a flexible and robust connectivity layer which is crucial in realizing the vision of Industry 4.0 and beyond \cite{TI_PIEEE}. \textcolor{black}{Private 5G simplifies conventional automation system hierarchies by providing interconnectivity on a wider and more fine-grained scale.  }

\textbf{Warehouse Operations} --
Mobile platforms like mobile robots and automated guided vehicles (AGVs) are  gaining popularity for various warehouse applications. Connectivity between a warehouse management system and mobile robots requires high reliability, very low and bounded latency and high scalability. 
With Private 5G, unprecedented opportunities arise for warehouse operations including  image-based or video-based guidance control of mobile robots, synchronized action between a fleet of mobile robots and low-cost remote control of mobile robots with minimal sensory capabilities.

\textbf{Utility Networks} -- Utility companies worldwide are rolling out smart metering networks for electricity, gas, water, etc. Private 5G provides the opportunity of deploying private smart metering networks for massive and secure data collection. Moreover, it offers the possibility of real-time demand/response management.  

\textbf{Industrial Remote Operation} -- Industrial sites like power plants, mines, building construction, oil/gas platforms and harbors can be extremely hazardous environments exposing personnel to a variety of risks. Remote operation allows humans to operate machinery/equipment with an increased level of safety and efficiency. It also provides a number of economic benefits like reduction of on-site workforce. 5G fulfils the latency requirements of real-time interaction. Private 5G provides the opportunity of industrial remote operation at industrial sites. Some of the key use-cases include remote operation of robotic equipment for nuclear decommissioning, remote crane operation at ports and harbors, and remote operation of construction and mining machinery.

\textbf{Mining Operations} -- Mines are typically located in remote locations where access to public networks is not always possible.  Moreover, mines require reliable coverage both inside and outside the mining sites. Private 4G has been deployed for mining operations in various parts of the world. For example, Telstra\footnote{https://tinyurl.com/wofv2qs} has deployed a private 4G network at a goldmine in Papua New Guinea.   Private 5G provides a promising opportunity to conduct mission-critical mining operations with greater safety and automation. It also connects the mining workforce with greater flexibility by services like push-to-talk and push-to-video. 

\textbf{Railway Networks} -- Overground/Underground railway transport networks require a mix of critical communication services for smooth operations. One such service is train radio which requires secure critical voice communication between the driver and the signaling controller. Another service is signaling communication between train and track side, e.g., in the case of communications-based train control (CBTC). Both services require very high system reliability and low latency. Private 5G provides an attractive solution in running a mix of critical services over a single technology. Moreover, it also enables enhanced safety services like train-to-track CCTV.

%
%
%
%
%
%

%
%


\section{Spectrum Opportunities for Private 5G}
Private 5G networks can be deployed in three different types of radio spectrum. 

\subsection{Licensed Spectrum}
Similar to public cellular networks, private 5G networks can be deployed in the licensed spectrum. Operation in licensed spectrum  provides greater certainty of performance with little risk of interference. Such operation is particularly attractive for mobile network operators (MNOs) in deploying private 5G networks. MNOs can  dedicate a portion of licensed spectrum for private network operation in a specific geographical area such as an industrial site. Regional regulatory bodies can also allocate spectrum for industrial networks.

\subsection{Unlicensed Spectrum}
\textcolor{black}{Another option for deployment of private 5G networks is the unlicensed spectrum, e.g., in the 2.4 GHz band,  the 5 GHz band or the recently-opened 6 GHz band \cite{FCC_report}.} These spectrum bands are used by Wi-Fi, Bluetooth, ZigBee, and various other technologies, and are inherently open for shared usage. Operation in licensed spectrum has received significant attention in context of 4G-LTE networks.  There are two main scenarios for operation of private 5G networks in unlicensed bands.

\begin{itemize}
\item \textbf{Standalone Unlicensed Operation} -- In this case, private 5G networks operate entirely in the unlicensed spectrum.  Unlicensed operation of 5G-NR is under investigation within 3GPP \cite{3gpp_38_889}. Such operation is particularly attractive for non-MNOs as private 5G networks can be deployed with no dependency on licensed spectrum. Its 4G counterpart is Multefire which supports unlicensed operation of LTE. Multefire implements a listen-before-talk (LBT) procedure to efficiently co-exist with other spectrum users in the same band. \textcolor{black}{Standalone unlicensed deployments are more appropriate for non-critical use-cases.  }
\item \textbf{Licensed Anchor Operation} -- This is similar to the licensed-assisted access (LAA) operation of LTE. In this case, operation in unlicensed band is supplemental to operation in the licensed band, i.e., unlicensed spectrum is aggregated with the licensed spectrum. 
\textcolor{black}{Such operation is particularly attractive for operator-deployed private networks seeking extra capacity. }
\end{itemize}


\subsection{Shared Licensed Spectrum}
The third option for private 5G deployment is the shared licensed spectrum. Operation in shared licensed spectrum opens a whole new range of possibilities, especially for non-MNOs. \textcolor{black}{Prominent examples of the shared licensed spectrum include the 3.5 GHz citizen broadband radio service (CBRS) band in the U.S., the 3.7 - 3.8 GHz band in Germany and the 3.8 - 4.2 GHz band in the U.K.  }

Unlike unlicensed spectrum, coordinated and dynamic spectrum access paradigms are emerging for  the shared spectrum which provide guarantees on interference-free operation similar to the licensed bands. One example is the three-tier citizens broadband radio service (CBRS) sharing model in the U.S. In addition to the incumbents, two types of spectrum users have been introduced: \emph{priority access license} (PAL) users and \emph{general authorized access} (GAA) users. The PAL users are licensed and must be protected from interference caused by other PAL users and GAA users. The GAA users are license-exempt and not entitled to protection from other tiers. Spectrum access for both PAL and GAA users is controlled by a dedicated spectrum access system (SAS).

The different private 5G operations models have been summarized in \tablename~\ref{oper_scenarios}.


\begin{table*}[]
\centering
\caption{\textcolor{black}{Private 5G Operational Scenarios} }
\begin{tabular}{llll}
\cline{1-4}
\toprule
\textbf{Functional Architecture  }                                                                                         & \textbf{Spectrum      }  & \textbf{Business Opportunity}              & \textbf{\textcolor{black}{Service Continuity}}                                                                                      \\
\cline{1-4}
\midrule
\multirow{3}{*}{Standalone Deployment}                                                                            & Licensed        & Deployed by MNOs                  & \multirow{3}{*}{\begin{tabular}[c]{@{}l@{}}Roaming agreements, Dual-radio, \\ N3IWF-based\end{tabular}} \\
                                                                                                                  & Unlicensed      & Deployed by MNOs or non-MNOs      &                                                                                                         \\
                                                                                                                  & Shared licensed & Deployed by non-MNOs              &                                                                                                         \\
                                                                                                                    \cline{1-4}
\multirow{3}{*}{\begin{tabular}[c]{@{}l@{}}Public-Private Shared\\ RAN Deployment\end{tabular}}                   & Licensed        & Deployed by MNOs                  & \multirow{3}{*}{\begin{tabular}[c]{@{}l@{}}Roaming agreements, Dual-radio, \\ N3IWF-based\end{tabular}} \\
                                                                                                                  & Unlicensed      & Deployed by MNOs or non-MNOs      &                                                                                                         \\
                                                                                                                  & Shared licensed & Deployed by non-MNOs              &                                                                                                         \\
                                                                                                                   \cline{1-4}
\multirow{2}{*}{\begin{tabular}[c]{@{}l@{}}Public-Private Shared RAN and\\ Control-plane Deployment\end{tabular}} & Licensed        & \multirow{2}{*}{Deployed by MNOs} & \multirow{2}{*}{\begin{tabular}[c]{@{}l@{}}Direct access to public \\ network services\end{tabular}}    \\
                                                                                                                  & Unlicensed       \\   
                                                                                                                                                                                                                      \cline{1-4}
\end{tabular}
\label{oper_scenarios}
\end{table*}

\section{Design Aspects and Key Challenges}

\subsection{Network Slicing for Private 5G}
\textcolor{black}{Network slicing is widely recognized as the key enabler for unlocking the potential of 5G for industry verticals. Network slicing subverts the one-size-fits-all approach of previous mobile/cellular generations. The fundamental principle of network slicing is to create multiple logical networks over a common physical infrastructure such that each logical network is tailored to the specific requirements of an application. Network slicing provides the capability of isolation from business, technical, functional and operational perspectives. Network slicing is particularly important for private 5G networks. It provides the means to create \emph{a network within a network} \cite{5G_ACIA_CI} to deliver certain services. For example, a private 5G network can be deployed as an isolated and dedicated slice inside a public 5G network. Slicing within a private 5G network provides a number of benefits. Owing to the increasingly heterogeneous nature of industrial communication,  private 5G deployments are expected to serve a wide range of industrial applications with different QoS requirements, ranging from conventional closed-loop control to emerging mobile robots. Network slicing is particularly attractive for co-existence of different services/applications within a private 5G network. Network slicing provides the capability of traffic isolation in an end-to-end manner which is important for providing strict performance guarantees (especially for critical industrial applications) in multi-service co-existence scenarios.  Network slicing also provides isolation in terms of computing, storage and networking resources. This ensures that the private 5G infrastructure is efficiently shared among different applications. Network slicing also provides the capability of optimizing a logical network based on slice requirements. This provides the opportunity for slice-specific resource management and security/privacy policies in a private 5G network.}


\textcolor{black}{Slicing of the RAN is an integral component of network slicing \cite{3gpp.38.300}. Broadly, there are two approaches for achieving network slicing in the RAN.} One approach is to achieve network slicing through   independent slicing of resources at individual base stations of a private network. However, such an approach would perform sub-optimally from a resource utilization perspective. It would also require modifications to standard base station protocols and procedures. \textcolor{black}{An alternative approach is to realize the slicing functionality in a gateway deployed inside a private 5G network. Such a gateway comes at higher levels in network hierarchy than base stations.  This approach provides many advantages. First, it enables network-wide slicing across the base stations of a private network which ensures efficient resource utilization. Second, it provides the capability of end-to-end resource slicing which is crucial in providing slice-specific performance guarantees. Third, it offers the possibility of independent slice customization at individual base stations. Finally, and most importantly, it hold minimum footprint for adoption in private 5G deployments as the gateway can be interfaced with third-party base stations.  } \textcolor{black}{Gateway-level slicing is aligned with the software-defined networking (SDN) paradigm for RAN \cite{softRAN}, \cite{net_slicing2}. One approach to realize gateway-level slicing is through an SDN controller for the RAN, e.g., based on the FlexRAN platform \cite{flexRAN}. The concept of gateway-based resource management for RAN has been investigated in literature. \textsf{NetShare} \cite{NetShare} and \textsf{AppRAN} \cite{AppRAN} provide gateway-based RAN sharing functionality. \textsf{5G-SliceR} \cite{slicing_patent} provides gateway-based slicing of wireless resources.  The role of gateways for network slicing is also under investigation in some standardization activities \cite{gateway_IETF}.    }


\textcolor{black}{Network slicing for 5G has been extensively studied \cite{netslice_survey1, netslice_survey2, netslice_survey3}. However, many important issues require further scrutiny, particularly in context of industrial networks and private deployments. To provide performance guarantees, end-to-end slicing encompassing joint slicing of radio/wireless, RAN and core network resources is required. Achieving end-to-end performance guarantees requires function-level isolation at different layers of the protocol stack. This entails functional decomposition of tightly coupled network functions in the RAN and the core network entities. Another aspect is modular and adaptive slice composition which requires functional abstraction of the network as well as abstraction of industrial assets, protocols and architectural hierarchies \cite{netslice_persp}.  A key part of end-to-end resource slicing with performance guarantees is the analysis of end-to-end properties of the network  \cite{netslice_ind}. Overall, achieving lightweight and minimal complexity gateway-level end-to-end network slicing in an important challenge for private 5G.     }

\subsection{Control-centric Radio Resource Allocation}
Network slicing provides the capability of application-specific customization. The term customization refers to optimization of resources allocated to a slice, in order to fulfil its service requirements. For example, in case of radio/wireless resources, conventional radio resource allocation techniques for human-centric applications like voice and video are not optimized for control-centric applications. Conventional techniques also treat uplink (device-to-network) and downlink (network-to-device) resource allocation as independent which is not suitable for control-centric applications, especially those involving control loops. Design of optimized radio resource allocation techniques for industrial control applications is yet another challenge for private 5G networks.

\subsection{Integration with TSN}
TSN is a set of standards under development within the IEEE 802.1 working group \cite{tsn} to improve security, reliability and real-time capabilities of \emph{standard} Ethernet. TSN provides guaranteed data delivery with deterministic and bounded low latency and extremely low data loss. TSN is widely recognized as the long-term replacement of proprietary mutually incompatible wired technologies in industrial domains. \textcolor{black}{However, wired connectivity is not feasible for many emerging industrial applications that are characterized by the requirements of mobility (e.g., automated guided vehicles) and flexibility (e.g., reconfigurable production lines).} \textcolor{black}{TSN is likely to co-exist with wireless systems. Integrated operation of TSN and 5G is crucial in achieving end-to-end deterministic connectivity in industrial systems.  }

\begin{figure}
\centering
\includegraphics[scale=0.32]{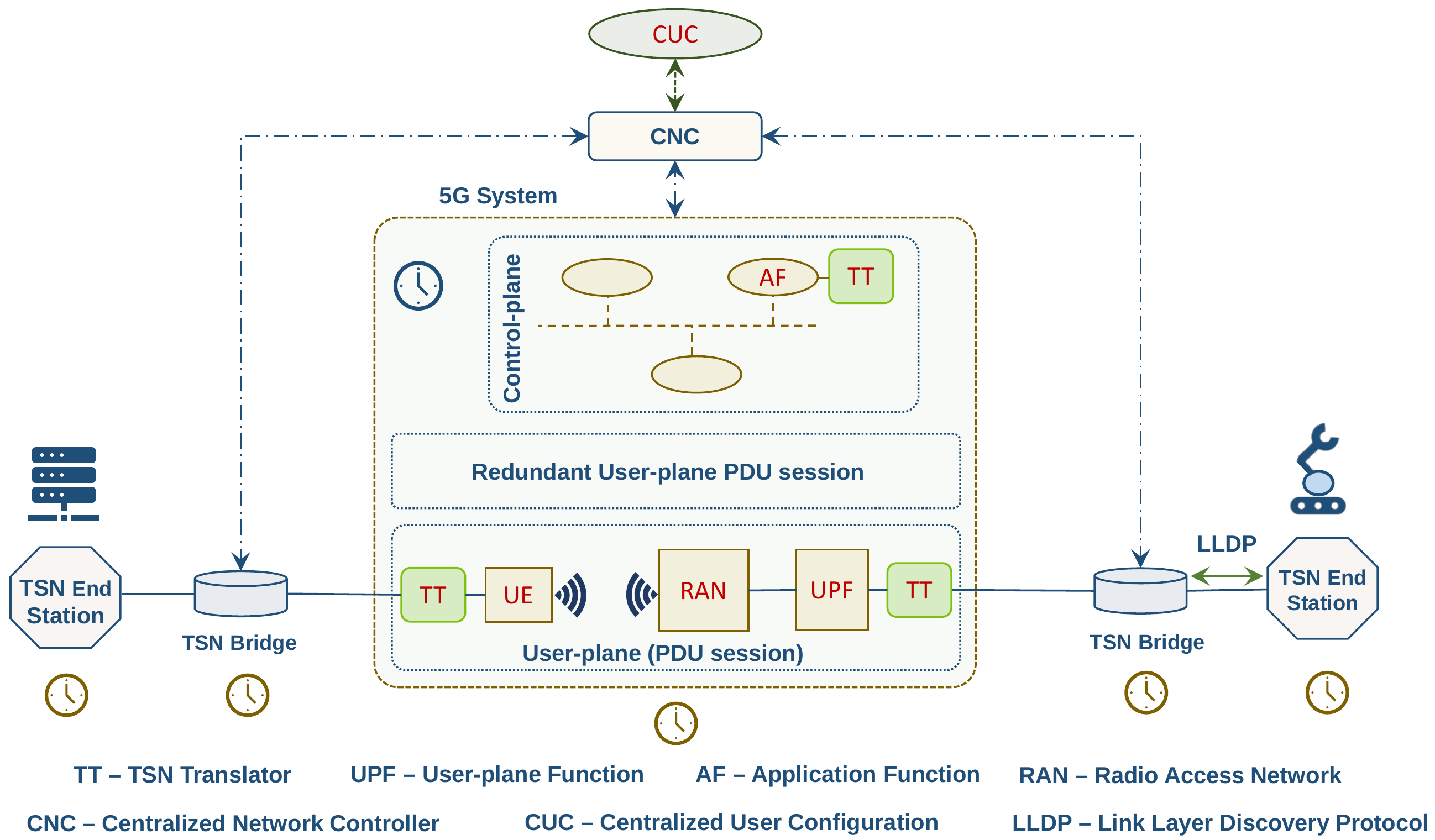}
\caption{\textcolor{black}{Tight integration of 5G and TSN systems based on the  bridge model. }}
\label{tsn_integ}
\end{figure}

\textcolor{black}{To realize converged operation, there must be tight (seamless) integration between 5G and TSN systems. 3GPP Release 16 is mainly focusing on the \emph{bridge model} \cite{3gpp_23_734} for integration which is illustrated in Fig. \ref{tsn_integ}. In this approach, the 5G system appears as a virtual TSN bridge or a black box to the TSN entities. The 5G system  handles TSN requests through its own QoS framework. It provides TSN ingress and egress ports via TSN translators. The primary advantage of the bridge model is that the 5G system does not need to support protocols and procedures that are part of the external TSN system. An alternative approach is the \emph{link model} where the 5G system appears  as a TSN link, i.e., as an Ethernet cable to the external network. This approach is challenging as the 5G system does not behave as an Ethernet cable, which leads to a fundamental mismatch between the capabilities of the two systems.}

\textcolor{black}{Realizing tight integration of 5G and TSN systems also creates various challenges. \textcolor{black}{In case of a 5G system bridge operating with a centralized TSN configuration, a TSN-compliant interface is required toward the centralized network configuration (CNC) entity in the TSN system}. Alignment of QoS between TSN and 5G systems is required to provide performance guarantees for TSN traffic.  Such alignment of QoS also necessitates resource management techniques for achieving TSN-like functionality over the 5G system. The 5G uRLLC framework provides the key enablers for realizing such resource management. For example, TSN frame replication and elimination for reliability  (FRER) can be achieved through recently standardized packet duplication techniques \cite{aijaz_pd_mag}. Moreover, network slicing techniques can isolate TSN application from other applications for meeting the requirements of TSN streams. Accurate time synchronization between TSN and 5G systems is another important challenge. Ongoing standardization activities \cite{3gpp_23_734} have addressed the key aspects of 5G-TSN time synchronization. \textcolor{black}{However, reference time indication techniques based on 5G protocols need to be closely examined under dynamically-scheduled and non-dedicated nature of signaling resources and potential sources of timing errors on the air-interface. } }



\section{Standardization and Open Innovation Ecosystem for Private 5G}
Historically, the industrial and wireless communities have worked in silos. This gap has been a major barrier in development and widespread adoption of industrial wireless technologies. In 2017, the German electrical and electronics manufacturing association, ZVEI, set up the Task Force 5G which has now expanded into \emph{5G Alliance for Connected Industries and Automation} (5G-ACIA)\footnote{https://www.5g-acia.org/}. This is a rather unprecedented development for bringing industrial and wireless communities together. Currently, 5G-ACIA has 50+ members that include all major stakeholders such as device manufacturers, chipset designers, operators, and infrastructure providers. The mission of 5G-ACIA is to bring the whole ecosystem together and to ensure that the requirements of the industrial domain are adequately addressed in 5G standardization and regulation. The development of 5G-ACIA is expected to accelerate adoption of 5G technology in industrial domains. 

On the other hand, 3GPP has been taking various initiatives in Release 16 which are focused on private 5G networks for industrial systems.  An early study \cite{3gpp.22.804} identified the requirements for 5G system for automation in vertical industries. 3GPP is also working on specific enhancements for operation of 5G in industrial domains such as support for TSN, time synchronization, Layer 2/3 optimization, unlicensed operation and enhanced quality-of-service (QoS) \cite{3gpp_38_825}.

Until recently, mobile/cellular technologies have been developed through closed innovation, and the stronghold of telecommunication vendors and infrastructure providers. 
Another unprecedented development is the emerging ecosystem of open innovation around 5G hardware and software. Recent trends toward softwarization in the networking industry has also influenced the development of 5G architecture. 3GPP Release 15 has adopted a service-based architecture with a modularized control-plane which is well-aligned with network function virtualization (NFV) principles. The open-RAN (O-RAN) alliance is focusing on developing an open RAN architecture for 5G on based the principles of openness and intelligence. The OpenRAN initiative of Telecom Infra Project (TIP)\footnote{https://telecominfraproject.com/} is focusing on fully programmable RAN solutions based on general-purpose vendor-neutral hardware and software-defined technology. The OpenAirInterface software alliance (OSA)\footnote{https://www.openairinterface.org/} is providing standard-compliant open source software stack for 4G/5G systems. \textcolor{black}{Mosaic5G\footnote{http://mosaic-5g.io/} is providing open source service platforms for 5G.  } The open innovation model for 5G is gaining traction. Numerous open source initiatives focusing on infrastructure, management, control, access, and core have emerged recently that are well-supported by industry. 

\textcolor{black}{The open innovation ecosystem is attractive for development of white box technology for private 5G. It minimizes dependency on vendors, infrastructure providers and MNOs. Non-MNOs and non-cellular companies can also tap into the value chain of private 5G by developing solutions using open source hardware and software. However, it also presents a number of challenges. Open source software and application programming interfaces (APIs) must be standardized for interoperability and widespread adoption. Devices based on open source stack and programmable hardware require validation and conformance testing. In context of industrial communication, product certification schemes must consider connectivity, interfacing, safety and security implications.  }

\section{Concluding Remarks} \label{sect_cr}
What 5G offers for industrial communication is well beyond the capabilities of any other wireless technology. Private 5G networks would be instrumental in unleashing the potential of 5G in the industrial sector.  Private (non-public) 5G  networks offer dependable industrial wireless communication and can be deployed for a wide range of industrial use-cases.  The opening of shared licensed spectrum and the emerging standardization and open innovation ecosystem would be the catalyst for private 5G deployments in industrial domains. While potentially opening new business opportunities for MNOs, private 5G empowers industrial players and non-MNOs to deploy their own networks. key design challenges for private 5G include lightweight end-to-end network slicing solutions, control-centric radio resource allocation techniques and seamless integration with TSN. The integration of TSN and private 5G would ultimately provide a single standardized industrial solution - something the industrial community has desired ever since the emergence of industrial networking. \textcolor{black}{Private 5G can unequivocally be foreseen as the future of industrial wireless over the next decade.} \textcolor{black}{Some of the key directions for future work include addressing the aforementioned design challenges, performance benchmarking of 5G against other wireless technologies in private deployments, coverage planning/optimization for private 5G networks, and public-private interworking.}







\bibliographystyle{IEEEtran}

\bibliography{IEEEabrv,Pvt5G}

\begin{thebibliography}{10}
\providecommand{\url}[1]{#1}
\csname url@samestyle\endcsname
\providecommand{\newblock}{\relax}
\providecommand{\bibinfo}[2]{#2}
\providecommand{\BIBentrySTDinterwordspacing}{\spaceskip=0pt\relax}
\providecommand{\BIBentryALTinterwordstretchfactor}{4}
\providecommand{\BIBentryALTinterwordspacing}{\spaceskip=\fontdimen2\font plus
\BIBentryALTinterwordstretchfactor\fontdimen3\font minus
  \fontdimen4\font\relax}
\providecommand{\BIBforeignlanguage}[2]{{%
\expandafter\ifx\csname l@#1\endcsname\relax
\typeout{** WARNING: IEEEtran.bst: No hyphenation pattern has been}%
\typeout{** loaded for the language `#1'. Using the pattern for}%
\typeout{** the default language instead.}%
\else
\language=\csname l@#1\endcsname
\fi
#2}}
\providecommand{\BIBdecl}{\relax}
\BIBdecl

\bibitem{marshall_ind}
P.~S. Marshall, ``{A Comprehensive Guide to Industrial Networks: Part 1},''
  \emph{Sensor Mag.}, vol.~18, no.~6, June 2001.

\bibitem{emer_ind_net}
J.~R. Moyne and D.~M. Tilbury, ``{The Emergence of Industrial Control Networks
  for Manufacturing Control, Diagnostics, and Safety Data},'' \emph{Proc.
  IEEE}, vol.~95, no.~1, pp. 29--47, January 2007.

\bibitem{ind4}
R.~Drath and A.~Horch, ``{Industrie 4.0: Hit or Hype?}'' \emph{IEEE Ind.
  Electron. Mag.}, vol.~8, no.~2, pp. 56--48, June 2014.

\bibitem{ind_internet}
J.-Q. Li \emph{et~al.}, ``{Industrial Internet: A Survey on the Enabling
  Technologies, Applications, and Challenges},'' \emph{IEEE Commun. Surveys
  Tuts.}, vol.~19, no.~3, pp. 1504--1526, Third Quarter 2017.

\bibitem{TI_PIEEE}
A.~Aijaz and M.~Sooriyabandara, ``{The Tactile Internet for Industries: A
  Review},'' \emph{Proc. IEEE}, vol. 107, no.~2, pp. 414--435, 2018.

\bibitem{wless_ind}
A.~Willig, K.~Matheus, and A.~Wolisz, ``{Wireless Technology in Industrial
  Networks},'' \emph{Proc. IEEE}, vol.~93, no.~6, pp. 1130--1151, June 2005.

\bibitem{fieldbus}
J.-P. Thomesse, ``{Fieldbus Technology in Industrial Automation},'' \emph{Proc.
  IEEE}, vol.~93, no.~6, pp. 1073--1101, June 2005.

\bibitem{ind_ethernet}
J.-D. Decotignie, ``{Ethernet-Based Real-Time and Industrial Communications},''
  \emph{Proc. IEEE}, vol.~93, no.~6, pp. 1102--1117, June 2005.

\bibitem{gen_field_net}
T.~Sauter, ``{The Three Generations of Field-Level Networks -- Evolution and
  Compatibility Issues},'' \emph{IEEE Trans. Ind. Electron.}, vol.~57, no.~11,
  pp. 3585--3595, November 2010.

\bibitem{fut_ind_comm}
M.~Wollschlaeger, T.~Sauter, and J.~Jasperneite, ``{The Future of Industrial
  Communication: Automation Networks in the Era of the Internet of Things and
  Industry 4.0},'' \emph{IEEE Ind. Electron. Mag.}, vol.~11, no.~1, pp. 17--27,
  March 2017.

\bibitem{3gpp.22.804}
\BIBentryALTinterwordspacing
3GPP, ``{Study on Communication for Automation in Vertical Domains},'' {3rd
  Generation Partnership Project (3GPP)}, TR {22.804}, December 2017, {v1.0}.
  [Online]. Available:
  \url{http://www.3gpp.org/ftp//Specs/archive/22_series/22.804/}
\BIBentrySTDinterwordspacing

\bibitem{ITU_2083}
\BIBentryALTinterwordspacing
ITU-R, ``{IMT Vision – Framework and Overall Objectives of the Future
  Development of IMT for 2020 and Beyond},'' {International Telecommunication
  Union (ITU)}, Recommendation ITU-R {M.2083-0}, Sept. 2015. [Online].
  Available:
  \url{https://www.itu.int/dms_pubrec/itu-r/rec/m/R-REC-M.2083-0-201509-I!!PDF-E.pdf}
\BIBentrySTDinterwordspacing

\bibitem{3gpp_22_872}
\BIBentryALTinterwordspacing
3GPP, ``{Study on Positioning Use cases Stage 1 (Release 16)},'' {3rd
  Generation Partnership Project (3GPP)}, TR {22.872}, Sept. 2018, {v16.1.0}.
  [Online]. Available:
  \url{https://www.3gpp.org/ftp/Specs/archive/22_series/22.872/}
\BIBentrySTDinterwordspacing

\bibitem{aijaz_IES}
\BIBentryALTinterwordspacing
A.~{Aijaz}, ``{High-Performance Industrial Wireless: Achieving Reliable and
  Deterministic Connectivity over IEEE 802.11 WLANs},'' \emph{IEEE Open Journal
  of the Industrial Electronics Society}, 2020. [Online]. Available:
  \url{https://arxiv.org/abs/2003.10188}
\BIBentrySTDinterwordspacing

\bibitem{3gpp_23_251}
\BIBentryALTinterwordspacing
3GPP, ``{Network sharing; Architecture and functional description},'' {3rd
  Generation Partnership Project (3GPP)}, TS {23.251}, Sept. 2018, {v15.1.0}.
  [Online]. Available:
  \url{https://www.3gpp.org/ftp/Specs/archive/23_series/23.251/}
\BIBentrySTDinterwordspacing

\bibitem{5G_ACIA_NPN}
5G-ACIA, ``{5G Non-Public Networks for Industrial Scenarios},'' {5G Alliance
  for Connected Industries and Automation}, White paper, July 2019.

\bibitem{3gpp_23_734}
\BIBentryALTinterwordspacing
3GPP, ``{Study on Enhancement of 5G System (5GS) for Vertical and Local Area
  Network (LAN) Services (Release 16)},'' {3rd Generation Partnership Project
  (3GPP)}, TR {23.734}, June 2019, {v16.2.0}. [Online]. Available:
  \url{https://www.3gpp.org/ftp/Specs/archive/23_series/23.734/}
\BIBentrySTDinterwordspacing

\bibitem{harbor_pvt}
\BIBentryALTinterwordspacing
{Harbor Research}, ``{The Private LTE Opportunity for Industrial and Commercial
  IoT},'' Report, July 2017. [Online]. Available:
  \url{https://harborresearch.com/private-lte-opportunity/}
\BIBentrySTDinterwordspacing

\bibitem{FCC_report}
\BIBentryALTinterwordspacing
{FCC}, ``{Unlicensed Use of the 6 GHz Band},'' Report {ET Docket No. 18-295},
  April 2020. [Online]. Available:
  \url{https://docs.fcc.gov/public/attachments/DOC-363490A1.pdf}
\BIBentrySTDinterwordspacing

\bibitem{3gpp_38_889}
\BIBentryALTinterwordspacing
3GPP, ``{Study on NR-based Access to Unlicensed Spectrum},'' {3rd Generation
  Partnership Project (3GPP)}, TR {38.889}, Dec. 2018, {v16.0.0}. [Online].
  Available: \url{https://www.3gpp.org/ftp/Specs/archive/38_series/38.889/}
\BIBentrySTDinterwordspacing

\bibitem{5G_ACIA_CI}
5G-ACIA, ``{5G for Connected Industries and Automation},'' {5G Alliance for
  Connected Industries and Automation}, White paper, Feb. 2019.

\bibitem{3gpp.38.300}
\BIBentryALTinterwordspacing
3GPP, ``{NR and NG-RAN Overall Description},'' {3rd Generation Partnership
  Project (3GPP)}, TS {38.300}, Sept. 2017, {v1.0}. [Online]. Available:
  \url{http://www.3gpp.org/ftp/Specs/archive/38_series/38.300/}
\BIBentrySTDinterwordspacing

\bibitem{softRAN}
A.~Gudipati, D.~Perry, L.~E. Li, and S.~Katti, ``{SoftRAN: Software Defined
  Radio Access Network},'' in \emph{ACM SIGCOMM Workshop on Hot Topics in
  Software Defined Networking (HotSDN)}, 2013, pp. 25--30.

\bibitem{net_slicing2}
J.~Ordonez-Lucena \emph{et~al.}, ``{Network Slicing for 5G with SDN/NFV:
  Concepts, Architectures, and Challenges},'' \emph{IEEE Commun. Mag.},
  vol.~55, no.~5, pp. 80--87, May 2017.

\bibitem{flexRAN}
X.~Foukas \emph{et~al.}, ``{FlexRAN: A Flexible and Programmable Platform for
  Software-Defined Radio Access Networks},'' in \emph{ACM International
  Conference on Emerging Networking Experiments and Technologies (CoNEXT)},
  2016, p. 427–441.

\bibitem{NetShare}
R.~Mahindra, M.~Khojastepour, H.~Zhang, and S.~Rangarajan, ``{Radio Access
  Network Sharing in Cellular Networks},'' in \emph{IEEE International
  Conference on Network Protocols (ICNP)}, Oct 2013, pp. 1--10.

\bibitem{AppRAN}
J.~He and W.~Song, ``{AppRAN: Application-oriented Radio Access Network Sharing
  in Mobile Networks},'' in \emph{IEEE International Conference on
  Communications (ICC)}, June 2015, pp. 3788--3794.

\bibitem{slicing_patent}
\BIBentryALTinterwordspacing
A.~Aijaz, ``{Radio Resource Slicing in a Radio Access Network},'' 2017, {US}
  Patent App. 15/441564. [Online]. Available:
  \url{https://patents.google.com/patent/US10264461B2/}
\BIBentrySTDinterwordspacing

\bibitem{gateway_IETF}
\BIBentryALTinterwordspacing
S.~Homma \emph{et~al.}, ``{Gateway Function for Network Slicing},'' {Internet
  Engineering Task Force}, Internet-Draft, March 2020. [Online]. Available:
  \url{https://tools.ietf.org/html/draft-homma-rtgwg-slice-gateway-02}
\BIBentrySTDinterwordspacing

\bibitem{netslice_survey1}
X.~{Foukas}, G.~{Patounas}, A.~{Elmokashfi}, and M.~K. {Marina}, ``{Network
  Slicing in 5G: Survey and Challenges},'' \emph{IEEE Commun. Mag.}, vol.~55,
  no.~5, pp. 94--100, 2017.

\bibitem{netslice_survey2}
A.~{Kaloxylos}, ``{A Survey and an Analysis of Network Slicing in 5G
  Networks},'' \emph{IEEE Commun. Standards Mag.}, vol.~2, no.~1, pp. 60--65,
  2018.

\bibitem{netslice_survey3}
I.~{Afolabi} \emph{et~al.}, ``{Network Slicing and Softwarization: A Survey on
  Principles, Enabling Technologies, and Solutions},'' \emph{IEEE Commun.
  Surveys Tuts.}, vol.~20, no.~3, pp. 2429--2453, 2018.

\bibitem{netslice_persp}
F.~{Ansah}, M.~{Majumder}, H.~{de Meer}, and J.~{Jasperneite}, ``{Network
  Slicing : An Industry Perspective},'' in \emph{IEEE International Conference
  on Emerging Technologies and Factory Automation (ETFA)}, 2019, pp.
  1367--1370.

\bibitem{netslice_ind}
A.~E. {Kalør} \emph{et~al.}, ``{Network Slicing in Industry 4.0 Applications:
  Abstraction Methods and End-to-End Analysis},'' \emph{IEEE Trans. Ind.
  Informat.}, vol.~14, no.~12, pp. 5419--5427, 2018.

\bibitem{tsn}
\BIBentryALTinterwordspacing
``{IEEE 802.1 Time-Sensitive Networking}.'' [Online]. Available:
  \url{http://www.ieee802.org/1/pages/tsn.html}
\BIBentrySTDinterwordspacing

\bibitem{aijaz_pd_mag}
A.~{Aijaz}, ``{Packet Duplication in Dual Connectivity Enabled 5G Wireless
  Networks: Overview and Challenges},'' \emph{IEEE Commun. Standards Mag.},
  vol.~3, no.~3, pp. 20--28, Sept. 2019.

\bibitem{3gpp_38_825}
\BIBentryALTinterwordspacing
3GPP, ``{Study on NR Industrial Internet of Things (IoT); (Release 16)},'' {3rd
  Generation Partnership Project (3GPP)}, TR {38.825}, March 2019, {v16.0.0}.
  [Online]. Available:
  \url{https://www.3gpp.org/ftp/Specs/archive/38_series/38.825/}
\BIBentrySTDinterwordspacing

\end{thebibliography}
%

\begin{IEEEbiographynophoto}{Adnan Aijaz}
(M'14--SM'18) studied telecommunications engineering at King's College London, United Kingdom, where he received a Ph.D. degree in 2014 for research in wireless networks. He is a Principal Research Engineer at  Bristol Research and Innovation Laboratory of Toshiba Corporation. His recent research topics include industrial communication systems and automation networks, cyber-physical systems,  next generation Wi-Fi and cellular (5G and beyond) technologies, and robotics and autonomous systems.  He has been contributing to various national and international research projects and standardization activities related to 5G and industrial communication. 
\end{IEEEbiographynophoto}

\end{document}